\documentclass[ preprint, amsmath, amssymb, prc ]{revtex4-1}
\usepackage{ graphicx, dcolumn, bm }

\begin{document}

\preprint{APS/123-QED}
\title{Impact of New $\beta$-decay Half-lives on $r$-process Nucleosynthesis}

\author{Nobuya Nishimura$^{1,\;2\;}$}
\email{nobuya.nishimura@unibas.ch}
\author{Toshitaka Kajino$^{3,\;4\;}$}
\author{Grant J. Mathews$^{5}$}
\author{\\Shunji Nishimura$^{6}$}
\author{Toshio Suzuki$^{7}$}

\affiliation{$^{1}$ Department of Physics, University of Basel, 4056 Basel, Switzerland}
\affiliation{$^{2}$ GSI, Helmholtzzentrum f\"ur Schwerionenforschung GmbH, 64291 Darmstadt, Germany}
\affiliation{$^{3}$ Division of Theoretical Astronomy, NAOJ, 181-8588 Mitaka, Japan}
\affiliation{$^{4}$ Department of Astronomy, The University of Tokyo, 113-033 Tokyo, Japan}
\affiliation{$^{5}$ Center for Astrophysics, University of Notre Dame, Notre Dame, IN 46556, USA}
\affiliation{$^{6}$ RIKEN Nishina Center, Wako, Saitama 351-0198, Japan}
\affiliation{$^{7}$ Department of Physics, Nihon University, 156-8550 Tokyo, Japan}

\date{\today}

\begin{abstract}
We investigate the effects of newly measured $\beta$-decay half-lives on $r$-process nucleosynthesis.
These new rates were determined by recent experiments
at the radioactive isotope beam factory facility in the RIKEN Nishina Center.
We adopt an $r$-process nucleosynthesis environment based on
a magnetohydrodynamic supernova explosion model
that includes strong magnetic fields and rapid rotation of the progenitor.
A number of the new $\beta$-decay rates are for nuclei on or near the $r$-process path,
and hence, they affect the nucleosynthesis yields and timescale of the $r$-process.
The main effect of the newly measured $\beta$-decay half-lives 
is an enhancement in the calculated abundance of isotopes with mass number
$A = 110$ -- $120$ relative to calculated abundances based upon $\beta$-decay rates estimated
with the finite-range droplet mass model.
This effect slightly alleviates, but does not fully explain, the tendency of
$r$-process models to underproduce isotopes with $A = 110$ -- $120$
compared to the solar-system $r$-process abundances.
\begin{description}
	\item[PACS numbers] 23.40.-s, 25.30.-c, 26.30.Hj, 26.50.+x, 97.60.Bw
\end{description}
\end{abstract}
\maketitle

%%%%%%%%%%%%%%%%%%%%%%%%%%%%%%%%%%%%%%%%%%%%%%%%%%%%%%%%%%%%%%%%%%%%%%%%%%%%%%%%%%%%%%%%%%%%%%
%     Section 1                                                                              %
%%%%%%%%%%%%%%%%%%%%%%%%%%%%%%%%%%%%%%%%%%%%%%%%%%%%%%%%%%%%%%%%%%%%%%%%%%%%%%%%%%%%%%%%%%%%%%

{\it Introduction.}
Rapid neutron-capture ($r$)-process nucleosynthesis
is responsible for the origin of approximately
half of the elements heavier than iron and
is the only means of producing the naturally occurring
radioactive heavy actinide elements such as Th and U.
In spite of more than a half century of study and observational progress
\citep[e.g.,][]{Sneden:etal:2008}, however,
the astrophysical sites for $r$-process nucleosynthesis
have still not been unambiguously identified
(for recent reviews see, e.g., \cite{Arnould:etal:2007, Thielemann:etal:2011}).
Although many candidate sites have been proposed and supernovae
appear to be well suited as the $r$-process site \citep{MathewsCowan:1990},
up till now there has been no consensus as to the correct astrophysical model.

Notwithstanding the difficulties in finding a suitable astronomical environment,
the physical conditions for the $r$-process are well constrained \cite{Burbidge:etal:1957}.
It is evident that the $r$-process occurs via a sequence of near equilibrium rapid neutron captures
and photo-neutron emission reactions far on the neutron-rich side of stability.
This equilibrium is established with a maximum abundance strongly peaked
on one or two isotopes far from stability.
The relative abundance of $r$-process elements is then determined
by the relative $\beta$-decay rates along this $r$-process path.,
i.e., slower $\beta$-decay lifetimes result in higher abundances.
At least part of the reason for the difficulty in finding the astrophysical site
for the $r$-process stems from the fact that it lies so far
from the region of stable isotopes
where there is little experimental data on $\beta$-decay rates.

In this context, it is of particular interest
that $\beta$-decay half-lives of 38 neutron-rich isotopes including
$^{100}$Kr, $^{103 - 105}$Sr, $^{106 - 108}$Y, $^{108 - 110}$Zr,
$^{111, 112}$Nb, $^{112 - 115}$Mo, and $^{116, 117}$Tc
have been measured \citep{Nishimura:etal:2011} at the recently commissioned
radioactive isotope beam factory (RIBF) facility at the RIKEN Nishina Center.
A secondary beam, containing the neutron-rich nuclei of interest,
was produced by inflight fission of a $345$ MeV/nucleon $^{238}$U beam in a Be target.
The nuclei in the secondary beam were identified on an event-by-event basis.
Their atomic numbers were determined from the energy loss in an ionization chamber,
and the charge to mass ratio was determined by combining the projectile time-of-flight
and magnetic rigidity measurements.
The nuclei were then implanted in a silicon-strip detector
where the $\beta$-decay lifetimes could be measured.
Most of the measured lifetimes are an improvement on existing measurements
and a number of them \citep{Nishimura:etal:2011} were measured for the first time.

Many of these isotopes are near or directly on the $r$-process path.
As such, they are of particular interest as they determine
the $\beta$-flow toward the important $r$-process peak at $A=130$.
Thus, they regulate the ability of models for the $r$-process
to form heavier elements \citep{Otsuki:etal:2003}.
It is of particular interest, therefore,
to examine the impact of these new rates on $r$-process models.

The magnetohydrodynamic (MHD) supernova explosion model is of particular interest
in the present study as it provides a good fit to the observed $r$-process abundances
and avoids some of the problems \citep[][and references therein]{Otsuki:etal:2003}
associated with other paradigms
such as the neutrino driven supernova wind scenario \citep{Woosley:etal:1994}.
Robust $r$-process nucleosynthesis appears to be a natural consequence of
MHD mechanisms for core-collapse supernova explosions involving strong magnetic fields and rapid rotation.
These models are also consistent with the observed jets
and asymmetry of core-collapse supernovae.
The detailed hydrodynamic properties of the MHD supernova model
and the input nuclear physics are described below.

%%%%%%%%%%%%%%%%%%%%%%%%%%%%%%%%%%%%%%%%%%%%%%%%%%%%%%%%%%%%%%%%%%%%%%%%%%%%%%%%%%%%%%%%%%%%%%
%     Section 2                                                                              %
%%%%%%%%%%%%%%%%%%%%%%%%%%%%%%%%%%%%%%%%%%%%%%%%%%%%%%%%%%%%%%%%%%%%%%%%%%%%%%%%%%%%%%%%%%%%%%
{\it Magnetohydrodynamic Models.}
The main features of the MHD supernova model employed here
are described in Ref. \cite{Nishimura:etal:2006} and need not be repeated here.
However, we emphasize that although this calculation is performed
in the context of a particular $r$-process paradigm,
this model is selected because it matches the required conditions of the $r$-process environment
(timescale, neutron density, temperature, entropy, electron fraction, etc.).
Thus, the results presented here should be similar to those obtained
from more generic $r$-process models \citep[e.g.,][]{Otsuki:etal:2003}.

\begin{figure}[htbp]
	\begin{center}
    \includegraphics[width=0.6\hsize]{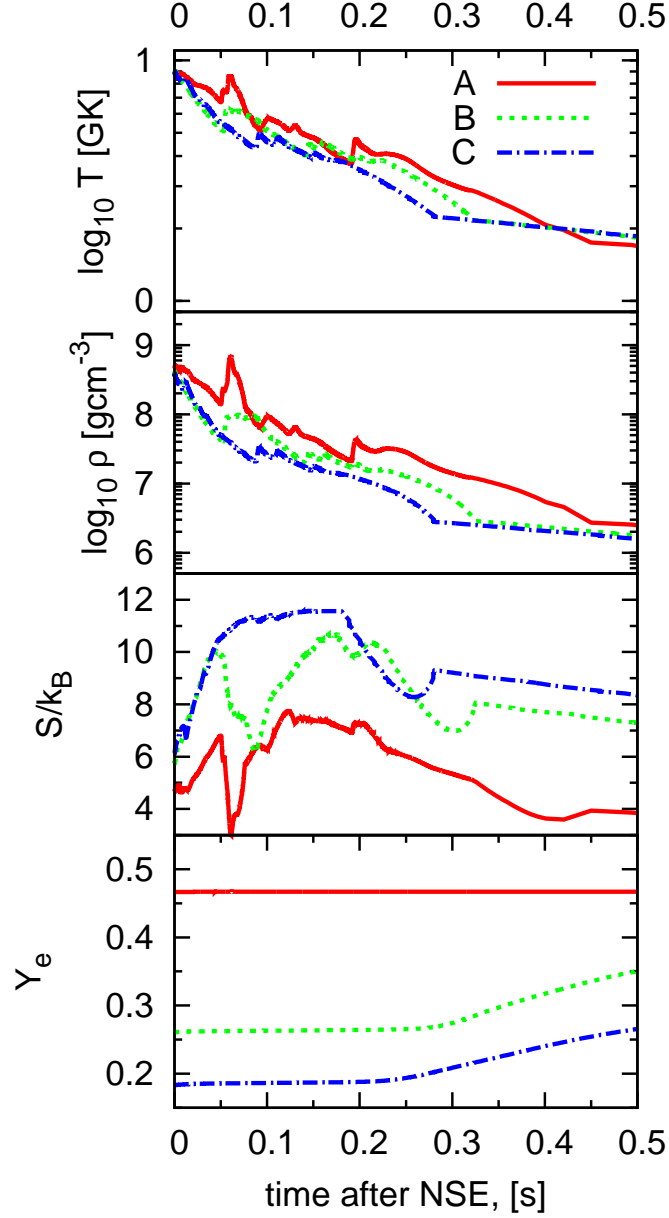}
    \caption{\label{traj}
    Thermodynamic properties for three representative tracer particles
    are shown as A ({\it red solid lines}), B ({\it green dotted lines}), and C ({\it blue dashed lines}).
    From top to bottom, each figure shows the time evolution of temperature $T$ in GK,
    density $\rho$ in g $\rm{cm}^{-3}$,
    entropy per baryon $S/k_{\rm{B}}$ in Boltzmann constant $k_{\rm{B}}$,
    and electron fraction $Y_{\rm{e}}$, respectively.
    The evolution of each variable is plotted for $0.5$ s from the time of the switch
    from NSE to a network calculation at $T = 9$ GK.}
	\end{center}
\end{figure}

We utilize the jet-like explosion (model 4 in \cite[][]{Nishimura:etal:2006})
based upon a two-dimensional magnetohydrodynamic simulation.
The ejecta were evolved with 23 tracers to describe the evolution
of thermodynamic state variables.
These were then post-processed to obtain the nucleosynthesis yields.
Figure \ref{traj} shows the evolution of entropy,
measured in units of the Boltzmann constant $k_{B}$,
and electron fraction, $Y_{\rm{e}}$, after nuclear statistical equilibrium (NSE)
is achieved for three representative tracer particles together with temperature and density.
Hydrodynamic quantities (e.g., entropy, temperature, and density, etc.)
are obtained from the  magnetohydrodynamic simulation,
whereas the electron fractions are deduced from the  nucleosynthesis calculation.
We switch from the NSE abundances to a nucleosynthesis network calculation
at $T = 9 \ \rm{GK}$ ($\sim 0.78 \ \rm{MeV}$).

%%%%%%%%%%%%%%%%%%%%%%%%%%%%%%%%%%%%%%%%%%%%%%%%%%%%%%%%%%%%%%%%%%%%%%%%%%%%%%%%%%%%%%%%%%%%%%
%     Section 3                                                                              %
%%%%%%%%%%%%%%%%%%%%%%%%%%%%%%%%%%%%%%%%%%%%%%%%%%%%%%%%%%%%%%%%%%%%%%%%%%%%%%%%%%%%%%%%%%%%%%

{\it Nuclear Reaction Network.}
The nuclear reaction network utilized for the nucleosynthesis simulations
has been described in detail elsewhere \cite{Nishimura:etal:2006,Nishimura:etal:2011b}
and need not be described in more detail here.
The network consists of more than $4000$ isotopes, including neutrons, protons,
and heavy isotopes with atomic number $Z \leq 100$
(for detail, see Table 1 in Ref. \cite{Nishimura:etal:2006}).
We also consider possible reactions related to the $r$-process
involving two- and three-body reactions or decay channels,
and we include electron capture as well as positron capture
and screening effects for all of the relevant charged particle reactions.
Experimentally determined masses \cite{AudiWapstra:1995} and reaction rates are adopted if available.
Otherwise, the theoretical predictions for nuclear masses, reaction rates,
and $\beta$-decays are obtained from
the finite-range droplet model (FRDM) \citep{Moller:etal:1995}.

% % % % % Table % % % % %
\begin{table}[b]
\caption{\label{tab-properties}
Reaction rates used in nuclear reaction networks}
\begin{ruledtabular}
\begin{tabular}{lll}
Network & $\beta$-decay half-lives & $(n,\gamma)$, $(\gamma,n)$ reaction \\
\hline
	FRDM  & REACLIB\footnotemark[1] & REACLIB \\
	RIBF  & REACLIB + RIKEN\footnotemark[2]   & REACLIB \\
	RIBF+ & REACLIB + RIKEN   & REACLIB + $Q_{\rm{+}}$ modified\footnotemark[3]
\footnotetext[1]{REACLIB : the REACLIB compilation from Ref. \cite{RauscherThielemann:2000}.}
\footnotetext[2]{RIKEN : experimental data given in Ref. \cite{Nishimura:etal:2011}.}
\footnotetext[2]{$Q_{+}$ modified : defined in Eq. (\ref{qvalue}).}
\end{tabular}
\end{ruledtabular}
\end{table}
% % % % % Table % % % % %

In the present study, we perform $r$-process calculations based upon
three different nuclear reaction networks as summarized in Table \ref{tab-properties}.
These are extensions of the basic network described above. 
One of the networks utilizes only the FRDM theoretical rates 
from the REACLIB compilation \cite{RauscherThielemann:2000}.
The other two (RIBF and RIBF+) utilize the new experimental $\beta$-decay half-lives
of 38 neutron-rich isotopes from \rm{Kr} to \rm{Tc}
and two versions of the theoretical FRDM $\beta$-decay rates for the other isotopes.
The  RIBF network replaces the FRDM rates with the new measured ones where possible.
The network (RIBF$+$) is based on the RIBF network and FRDM rates
with modified $Q$ values for $(n,\gamma)$ and reverse reactions given by:
\begin{equation}
	Q_{\rm{+}} =
	\begin{cases}
	Q - 0.3 ~\rm{[MeV]} &(\;\;97 \le A \le 103) \\
	Q + 0.5 ~\rm{[MeV]} &(104 \le A \le 107) \\
	Q + 1.0 ~\rm{[MeV]} &(108 \le A \le 115)
	\label{qvalue}
	\end{cases}
\end{equation}
where $Q$ is the theoretical $Q$ value (in MeV) obtained from the FRDM \citep{Moller:etal:1995}.
The motivation for these  modified $Q$ values is that they lead to a better fit
to the observed $\beta$-decay lifetimes away from stability.

\begin{figure}[htbp]
	\begin{center}
    \includegraphics[width=0.6\hsize]{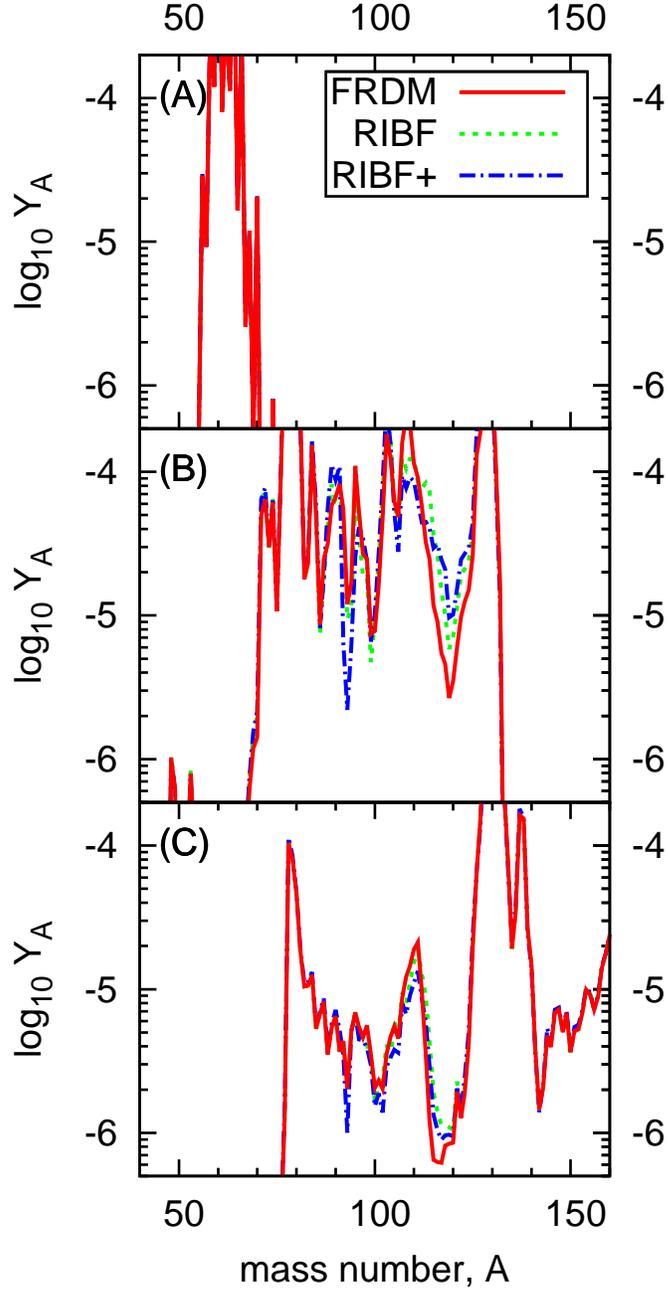}
    \caption{
    Upper, middle, and lower panels show final abundance distributions
	for the illustrative tracer particles A, B, and C from Figure \ref{traj}.
	Each panel shows three different lines corresponding
	to the different reaction networks, i.e., FRDM ({\it solid red lines}),
	RIBF ({\it dashed green lines}), and RIBF+ ({\it dotted blue lines}).
	In the case of tracer A, the three lines are indistinguishable
	because the abundance distribution does not reach mass numbers
	$A \sim 100$ -- $130$ where differences in the networks appear.}
\label{traj-abund}
	\end{center}
\end{figure}

%%%%%%%%%%%%%%%%%%%%%%%%%%%%%%%%%%%%%%%%%%%%%%%%%%%%%%%%%%%%%%%%%%%%%%%%%%%%%%%%%%%%%%%%%%%%%%
%     Section 4                                                                              %
%%%%%%%%%%%%%%%%%%%%%%%%%%%%%%%%%%%%%%%%%%%%%%%%%%%%%%%%%%%%%%%%%%%%%%%%%%%%%%%%%%%%%%%%%%%%%%
{\it Results.}
The nucleosynthesis calculations have been performed for
$23$ tracer particles from the MHD supernova model (e.g. Figure \ref{traj})
with the three different reaction networks listed in Table \ref{tab-properties}.
In particular, the underproduction of isotopes near $A = 120$
becomes slightly less pronounced relative to predictions based upon
the FRDM rates when the new measured rates are employed. 
Figure \ref{traj-abund} illustrates the final abundances obtained
from the three representative trajectories (A, B, and C) of Figure \ref{traj}
and the three reaction networks employed.
Note that as the $A \sim 130$ $r$-process peak is approached,
the new rates begin to make a difference in isotopes with $A \sim 90$ -- $120$.

\begin{figure*}[htbp]
	\begin{center}
		\includegraphics[width=\hsize]{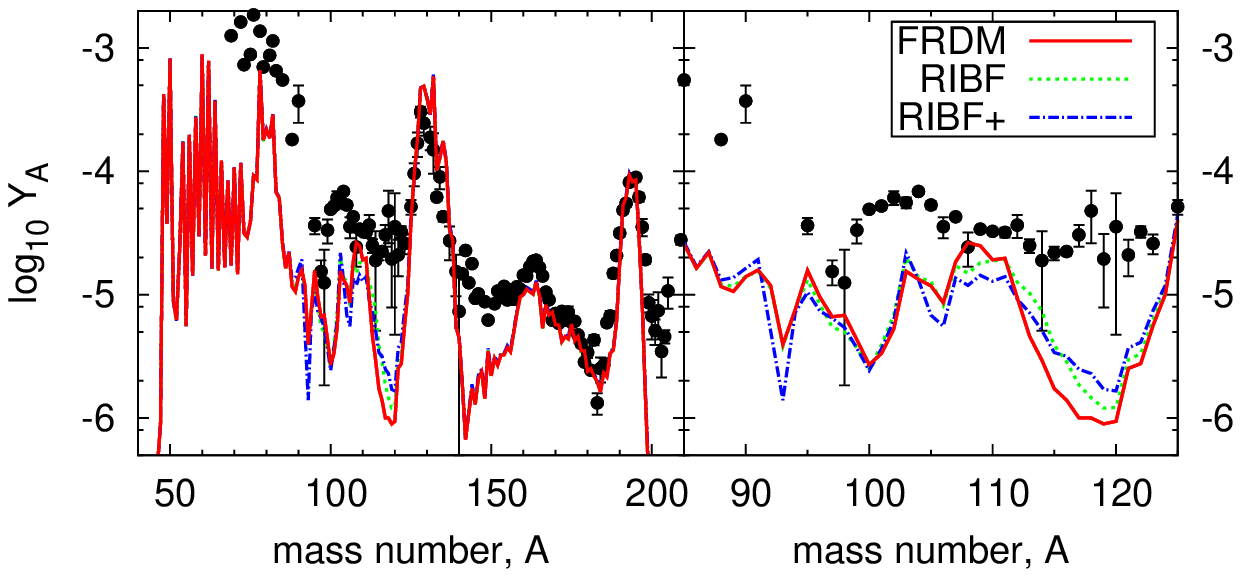}
		\caption{
		Integrated mass averaged total final abundance distributions
		of $r$-process elements from the adopted MHD supernova model
		(jet model 4 in Ref. \citep{Nishimura:etal:2006}).
		Red solid, green dotted, and blue dashed lines correspond
		to results from using the FRDM (standard), RIBF, and RIBF+ rates, respectively.
		Abundances of solar system $r$-elements \citep{Arlandini:etal:1999}
		are represented by black dots with error bars.} 
		\label{tot-abund}
	\end{center}
\end{figure*}

Figure \ref{tot-abund} shows the final integrated abundance distribution
from all of the trajectories.
These are compared to the solar system $r$-process abundance distribution
\cite{Arlandini:etal:1999}.
Here the effect of new rates becomes more apparent.
There has been a recurrent conundrum in the $r$-process models
in that they tend to underproduce nuclei with $A \sim 120$ \citep{Woosley:etal:1994}.
One hope has been that the newly measured $\beta$-decay rates
in this mass region might shift the $\beta$-flow equilibrium
thereby filling in the low abundances near $A \sim 120$.
Here, however, we see on Figure \ref{tot-abund} that the abundances
in the $A = 110$ -- $120$ region are only slightly enhanced.
Thus, although the new rates provide a little assistance
in enhancing the abundances near the valley,
they do not alleviate this problem.

This suggests that a further modification of the $r$-process paradigm is required.
One possibility is suggested by this study.
We note that the $r$-process flow moved farther away from stability
and proceeded faster in the RIBF+ network in which the ($n,\gamma$) Q values
were systematically enhanced for isotopes with $A = 104$ -- $115$.
This helped to fill in the abundances
in the higher mass region with $A \ge 115$ (see Figure \ref{tot-abund}).
Thus, it is important to measure the masses
and/or neutron separation energies in this region.
For example, if the strength of the nuclear closed shell near the neutron magic number $82$,
and $A = 120$ -- $140$ was systematically diminished,
this would prevent the $r$-process path from bypassing the A $\sim 120$ nuclei
as the $(n,\gamma)$ equilibrium shifts toward the more strongly bound nuclei
along $N=82$ the neutron closed shell.
Whatever the explanation for the filling of $A \sim 110$ -- $120$,
however, it is evident that the new $\beta$-decay rates
have provided new insight into the nucleosynthesis of the heavy nuclei
in the $r$-process.

At the very least, these results confirm that the final abundances
in the mass region of $A \sim 110-120$ are quite sensitive
to the $\beta$-decay half-lives of isotopes along the $r$-process path.
Indeed, the astrophysics of the production of nuclei
in the mass region of $A \sim 110-120$ is currently of considerable interest.
It is presently thought that the lighter heavy elements
with $A \le 120$ observed in some ultra-metal-poor stars,
may be the result of a new light-element primary process (LEPP) \cite[][]{Travaglio:etal:2014}.
Our results suggest that nucleosynthesis in the LEPP
should also be sensitive to the nuclear physics uncertainty
from $\beta$-decay rates in this region.
Hence, further studies on both the nuclear physics and astrophysics of
the synthesis of elements with $A \sim 110-120$ are warranted.

\begin{acknowledgments}
This study was supported in part by Grants-in-Aid
for Scientific Research of JSPS (19340074, 20244035, and 22540290),
JSPS Fellows (21.6817), Scientific Research on Innovative Area of MEXT (20105004),
the U.S. National Science Foundation Grant No. PHY-0855082,
and U.S. Department of Energy under Nuclear Theory Grant DE-FG02-95- ER40934.
\end{acknowledgments}

\end{document}